\documentclass[11pt,a4paper]{article}
\pdfoutput = 1
\usepackage{jheppub}
\usepackage{amsmath,bm}
 
\newcommand{\mi}{\mathrm{i}}

\newcommand{\as}{\alpha_{\mathrm{s}}}

\newcommand{\LA}{\mathrm{A}}
\newcommand{\LB}{\mathrm{B}}
\newcommand{\LE}{\mathrm{E}}
\newcommand{\LF}{\mathrm{F}}

\newcommand{\LR}{\mathrm{R}}

\newcommand{\LT}{\mathrm{T}}
\newcommand{\La}{\mathrm{a}}
\newcommand{\Lb}{\mathrm{b}}
\newcommand{\Lc}{\mathrm{c}}

\newcommand{\Lf}{\mathrm{f}}
\newcommand{\Lg}{\mathrm{g}}

\newcommand{\GeV}{\ \mathrm{GeV}}
\newcommand{\TeV}{\ \mathrm{TeV}}

\def\ket#1{\big|{#1}\big\rangle}
\def\bra#1{\big\langle{#1}\big|}
\def\brax#1{\big\langle{#1}}   

\def\sket#1{\big|{#1}\big)}

\title{Effects of subleading color in a parton shower}

\author[a]{Zolt\'an Nagy}

\author[b]{and Davison E.\ Soper}

\affiliation[a]{
DESY\\
Notkestrasse 85\\
22607 Hamburg, Germany
}

\affiliation[b]{
Institute of Theoretical Science\\
University of Oregon\\
Eugene, OR  97403-5203, USA
}

\emailAdd{Zoltan.Nagy@desy.de}
\emailAdd{soper@uoregon.edu}

\abstract{
Parton shower Monte Carlo event generators in which the shower evolves from hard splittings to soft splittings generally use the leading color (LC) approximation, which is the leading term in an expansion in powers of $1/N_\Lc^2$, where $N_\Lc = 3$ is the number of colors. In the parton shower event generator \textsc{Deductor}, we have introduced a more general approximation, the LC+ approximation, that includes some of the color suppressed contributions. In this paper, we explore the differences in results between the LC approximation and the LC+ approximation. Numerical comparisons suggest that, for simple observables, the LC approximation is quite accurate. We also argue that for gap-between-jets cross sections neither the LC approximation nor the LC+ approximation is adequate.
}

\preprint{DESY 14-250}
 
\begin{document}
\maketitle


\section{Introduction}
\label{sec:Intro}

Parton shower Monte Carlo event generators like \textsc{Pythia} \cite{Pythia}, \textsc{Herwig} \cite{Herwig}, and \textsc{Sherpa} \cite{Sherpa} represent the momenta and flavors carried by partons in a natural way \cite{EarlyPythia, Gottschalk, angleorderMW, angleorderEMW}. One can imagine that at any ``shower time'' $t$ in the development of the shower, there is a probability $\rho(t,\{p_\La,f_\La,p_\Lb,f_\Lb,p_1,f_1,\dots,p_m,f_m\})$ to have $m$ final-state partons plus two initial-state partons with the specified momenta $p_l$ and flavors $f_l$. What happens between time $t$ and $t + \Delta t$ is a change $\Delta\sket{\rho(t)}$ chosen at random according to physics based parton splitting probabilities. The evolution of $\sket{\rho(t)}$ is straightforwardly described in the language of classical statistical mechanics.

Spin and color do not fit easily into the event generator format because quantum interference between different spin and color states is important. For this reason, the natural language is that of quantum statistical mechanics. Then $\sket{\rho(t)}$ represents not just a probability distribution, but a density matrix in color and spin space.

There is a fairly straightforward formalism \cite{NSI} available to specify an evolution equation for $\sket{\rho(t)}$ that takes full account of color and spin. Unfortunately, we do not know how to implement the full evolution equation in a practical computer program that would constitute a parton shower event generator. If we use the leading color (LC) approximation and average over spins, we do get something practical \cite{NSII}. 

We have implemented the resulting shower evolution equation as a parton shower event generator, \textsc{Deductor} \cite{deductor, DeductorCode}. In ref.~\cite{deductor}, we presented results from \textsc{Deductor} using the leading color approximation. However, there is an improved version of the leading color approximation that is still practical: the LC+ approximation \cite{NScolor}. The LC+ approximation is built into \textsc{Deductor}. 

In this paper, we use the LC+ approximation to explore numerically the effect of color in a parton shower. We do not have a shower treatment with full color. It should be possible \cite{NScolor} to expand cross sections perturbatively in the difference between full color splitting functions and their LC+ approximations, but this possibility is not implemented in \textsc{Deductor}. Thus we are limited to looking at what we can discover with the LC+ approximation. From this investigation, two tentative conclusions emerge. 

First, comparing results for simple observables calculated with the LC and LC+ approximations, we find differences of order $1/N_\Lc^2 \approx 10\%$, but not larger differences. This is perhaps not a surprise, but real calculations were needed to reach this conclusion. 

Second, in events with two high $p_\LT$ jets with a large rapidity separation between them, we argue that there can be a substantial influence of color on the probability to have a gap between the two jets that contains no jets with transverse momentum above a cut $p_{\LT}^{\rm cut}$. Here, neither the LC approximation nor the LC+ approximation is adequate.

This paper is organized as follows. We provide some background information in section \ref{sec:Background}. Then in section \ref{sec:howmuchI} we investigate how much color suppression is typical in the color state as the shower develops. (The brief answer is that it is rare to have no color suppression.) In sections \ref{sec:jetcrosssection} and \ref{sec:NinJet}, we look at the difference between LC results and LC+ results for two simple observables: the one jet inclusive cross section and the distribution of the number of partons in a jet (which is not really physically observable, but is calculable with a fixed cutoff on the $k_\LT$ in parton splittings). In section \ref{sec:gapfraction}, we examine the influence of color on the cross section to have a gap between high $p_\LT$ jets. We present some conclusions in section \ref{sec:conclusions}. There is an appendix \ref{sec:LC} about how to switch from the LC+ approximation to the LC approximation part way into the shower. Another approach to color in a parton shower has been presented in the literature \cite{PlatzerSjodahl}. We compare this approach to ours in appendix \ref{sec:PlatzerSjodahl}

\section{Background}
\label{sec:Background}

Before exploring numerical results, we provide a little background about \textsc{Deductor} and about color and the LC+ approximation.

\subsection{Deductor}
\label{sec:Deductor}

A brief presentation of the parton shower event generator \textsc{Deductor} can be found in ref.~\cite{deductor}. Here, we simply mention some major features. First, \textsc{Deductor} generates a parton shower, but does not at present include a hadronization stage. Nor does it include an underlying event. Thus it is well suited as a tool for investigating the approximations in parton shower algorithms, but has some limitations for other uses. Second, \textsc{Deductor} averages over parton spins at each stage of splittings, in the same fashion as other parton shower event generators. There is a method for including spin that should be practical \cite{NSspin}, but we have not yet implemented it. Third, we argue in ref.~\cite{ShowerTime} that it is best to order splittings in a parton shower in order of decreasing values of the virtuality in the splitting divided by the energy of the mother parton. This rather non-standard choice is implemented in \textsc{Deductor}. Fourth, initial-state charm and bottom quarks have their proper masses in \textsc{Deductor} \cite{MassivePdfs}.

The parton splittings in \textsc{Deductor} are ordered from hardest to softest. Quantum interference is accounted for at least approximately because the shower is based on color dipoles \cite{dipolesG, dipolesGP}. Another approach is to order splittings in angles, from largest to smallest \cite{angleorderMW, angleorderEMW}. In this approach, one makes a rather severe approximation with respect to the dependence of the splitting probability on the azimuthal angle of the emission, but one gains accuracy with respect to color structure. This is the approach followed in \textsc{Herwig} \cite{Herwig}.

\subsection{The color density matrix}
\label{sec:ColorDensityMatrix}

The natural language for describing an evolving probability distribution in a parton shower is statistical mechanics. In order to include quantum color, we need {\em quantum} statistical mechanics. Thus we need a density operator that depends on the shower time $t$ and is an operator on the space of color states of a possibly large number of partons. The density operator has the form
\begin{equation}
\label{eq:rhodef}
\rho(\{p,f\}_m,t) 
= \sum_{\{c\}_m ,\{c'\}_m}\rho(\{p,f,c',c\}_m,t)\,\ket{\{c\}_m}\bra{\{c'\}_m}
\;\;.
\end{equation}
Here $\ket{\{c\}_m}$ and $\ket{\{c'\}_m}$ are standard \cite{ColorBasis} basis vectors for the quantum color space for $m$ final state partons plus two initial-state partons \cite{NSI, NScolor}. The color configuration $\{c\}_m$ of the ket state is, in general, different from the color configuration $\{c'\}_m$ of the bra state. Thus the function $\rho(\{p,f,c',c\}_m,t)$ depends on two sets of color indices. The density operator $\rho(\{p,f\}_m,t)$ can be regarded as a vector in the space of functions of $\{p,f\}_m$ with values in the space of operators on the quantum color space.

The color basis states are normalized to $\brax{\{c\}_m}\ket{\{c\}_m} = 1$ or to $\brax{\{c\}_m}\ket{\{c\}_m} \approx 1$ with very small corrections. They are not, however, generally orthogonal. However, when $\{c\}_m$ and $\{c'\}_m$ are different, one finds that $\brax{\{c'\}_m}\ket{\{c\}_m} = {\cal O}(1/N_\Lc^P)$ with $P \ge 1$. That is, the basis vectors are orthogonal in the $N_\Lc \to \infty$ limit.

At the end of a parton shower, one will want to measure something about the final state, using a color singlet measurement operator.\footnote{This leaves out hadronization. If we apply a hadronization model based on the formation of color strings, we do, in effect, apply a non-color-singlet measurement operator. We formulate how this can work in ref.~\cite{NScolor}.} This means that we take the trace of $\rho(\{p,f\}_m,t)$ in the color space, so that the contribution from color states $\{c',c\}_m$ is proportional to $\brax{\{c'\}_m}\ket{\{c\}_m}$. That is, the contribution from color states with $\{c'\}_m \ne \{c\}_m$ is suppressed by one or more factors of $1/N_\Lc$.

One notes, however, that states with $\{c'\}_m \ne \{c\}_m$ are a natural part of quantum chromodynamics (QCD). They appear even at the Born level of a hard scattering, where we normally keep track of them by using color ordered amplitudes. Even if we start with $\{c'\}_m = \{c\}_m$ states, splittings in a parton shower generate $\{c'\}_m \ne \{c\}_m$ states. 

\subsection{The color suppression index}
\label{sec:ColorSuppressionIndex}

In order to understand the systematics how many factors of $1/N_\Lc$ accompany states with $\{c'\}_m \ne \{c\}_m$, it is useful to use the color suppression index \cite{NScolor}. 

To do that, we first note a small technical complication. In a $g \to q + \bar q$ splitting we have a color matrix $t^a_{ij}t^a_{i'j'}$ in the color amplitude. We can use the Fierz identity,
\begin{equation}
t^a_{ij}t^a_{i'j'}
= \frac{1}{2}\, \delta_{ij'}\delta_{i'j} 
- \frac{1}{2N_\Lc}\,\delta_{ij}\delta_{i'j'}
\;,
\end{equation}
to write the result expanded in color basis states. At each $g \to q + \bar q$ splitting, shower evolution picks either the first, leading color, term or else the second, color suppressed, term. If the second term is chosen, further evolution uses the second color state, $\delta_{ij}\delta_{i'j'}$, and incorporates the factor $-1/(2N_\Lc)$ into the weight factor for the event. We let $p_\LE$ represent the number of times during the shower evolution that we pick up a $1/N_\Lc$ factor by using the second term in the Fierz identity. 

With that complication out of the way, we can define the color suppression power $P$ as the number of powers of $1/N_\Lc$ that appear at some stage of the shower coming from both the overlap $\brax{\{c'\}_m}\ket{\{c\}_m}$ of the color states, and the explicit $1/N_\Lc$ factors from $g \to q + \bar q$ splittings:
\begin{equation}
\label{eq:colorpower}
\left(\frac{1}{N_\Lc}\right)^{\!p_\LE}
\brax{\{c'\}_m}\ket{\{c\}_m}
= \frac{c_P(m)}{N_\Lc^{P(m)}}
\left\{
1 + {\cal O}\left(\frac{1}{N_\Lc}\right)
\right\}
\;.
\end{equation}

Now we define the color suppression index $I$. We simply use the group $U(N_\Lc)$ in place of the true color group $SU(N_\Lc)$ to calculate the color overlap
\begin{equation}
\label{eq:colorindex}
\left(\frac{1}{N_\Lc}\right)^{\!p_\LE}
\brax{\{c'\}_m}\ket{\{c\}_m}_{U(N_\Lc)}
= \frac{c_I(m)}{N_\Lc^{I(m)}}
\left\{
1 + {\cal O}\left(\frac{1}{N_\Lc}\right)
\right\}
\;.
\end{equation}

The color suppression index has two properties that make it quite useful. First, we always have
\begin{equation}
\label{eq:PgtI}
P(m) \ge I(m)
\;.
\end{equation}
That is, if we use $I(m)$ to estimate the amount of color suppression, we can never overestimate. Second, at each stage of the shower, the color suppression index either stays the same or else it increases:
\begin{equation}
\label{eq:Igrows}
I(m+1) \ge I(m)
\;.
\end{equation}
Thus we can think of $I$ as measuring color disorder, like entropy: it can never decrease as the shower evolves.

\subsection{The LC and LC+ approximations}
\label{sec:LCandLCP}

We have noted above that the hard scattering that leads to a parton shower has some contributions with $\{c'\}_m = \{c\}_m$ and thus with color suppression index $I = 0$.\footnote{If the hard scattering is $2 \to 2$ parton-parton scattering, then $m = 2$ at the start of the parton shower.} It also generally has some contributions with $\{c'\}_m \ne \{c\}_m$ and thus with color suppression index $I > 0$. This implies that the associated probability has at least $I$ powers of $1/N_\Lc$. As the shower develops, each parton splitting gives some terms that leave $I$ unchanged, but also some terms that increase $I$. Thus, inevitably, we generate some color configurations that come with high powers of $1/N_\Lc$ and thus low probabilities.

The contributions with $I > 0$ are also associated with calculational difficulties in a parton shower. For this reason, it is common to start at the hard scattering with $\{c'\}_m = \{c\}_m$, $I = 0$ states only and then to omit generating any $I > 0$ states. In practice, it is very easy to avoid increasing $I$ in parton splittings. In simplest terms, we count each gluon as carrying color $\bm{3}\times \bar{\bm{3}}$ instead of color $\bm{8}$. This gives us the leading color approximation.

We can also use the LC+ approximation \cite{NScolor}. Here we can start with $\{c'\}_m \ne \{c\}_m$, $I > 0$ states from the hard process. Then we can keep some of the terms with increasing $I$ that are generated by parton splittings. We do not keep all of the terms, so this is still an approximation. The approximation is exact for collinear splittings and it is exact for splittings that are simultaneously collinear and soft. It does not, however, keep everything that arises from wide angle soft splittings. We refer the reader to ref.~\cite{NScolor} for a full description.

With the LC+ approximation, we can eventually generate states with quite high values of the color suppression index $I$. Because these states come along with numerical weights, this has the effect of slowing down the numerical convergence of the program. That is, it takes more events to give the same statistical error. For this reason, we set a maximum value $(\Delta I)_{\rm max}$ for how much can be added to the color suppression index by the parton shower. When the amount of color suppression added by the shower, $\Delta I$, reaches $(\Delta I)_{\rm max}$, we switch off the LC+ approximation and start using a version of the LC approximation that is adapted to $\{c'\}_m \ne \{c\}_m$. We describe how to do this in appendix \ref{sec:LC}. For numerical results in this paper, we set $(\Delta I)_{\rm max} = 4$. Shower splitting either leaves $I$ unchanged or increases it by two. Thus we are omitting terms with $\Delta I = 6$, corresponding to omitted factors as large as $1/N_\Lc^6 \approx 10^{-3}$, which should allow for sufficient accuracy for any practical purpose. Of course, $(\Delta I)_{\rm max}$ is adjustable, so that we have an approximation with errors that can be controlled.

\subsection{Systematic improvement in color}
\label{sec:PertExpansion}

Let us look briefly at how the LC+ approximation works and how it can be improved \cite{NScolor}. With full color, shower evolution has the form 
\begin{equation}
\label{eq:rhofromU}
\sket{\rho(t)} = {\cal U}(t,t_0)\sket{\rho(t_0)}
\;,
\end{equation}
where $\sket{\rho(t)}$ represents the state of the system at shower time $t$ in a statistical ensemble of trials. The evolution operator ${\cal U}(t,t_0)$ obeys the evolution equation,
\begin{equation}
\label{eq:evolutionU}
\frac{d}{dt}\,{\cal U}(t,t_0)
= [{\cal H}_I(t) - {\cal V}(t)]\,{\cal U}(t,t_0)
\;.
\end{equation}
Here ${\cal H}_I(t)$ generates a parton splitting, while ${\cal V}(t)$ leaves unchanged the number of partons and their momenta. For the LC+ approximation, we approximate ${\cal U}(t,t_0)$ by ${\cal U}^{{\rm LC+}}(t,t_0)$ where
\begin{equation}
\label{eq:evolutionLCplusbis}
\frac{d}{dt}\,{\cal U}^{{\rm LC+}}(t,t_0)
= [{\cal H}^{{\rm LC+}}_I(t) - {\cal V}^{{\rm LC+}}(t)]\,
{\cal U}^{{\rm LC+}}(t,t_0)
\;\;.
\end{equation}
This differential equation can be solved iteratively in the form
\begin{equation}
\label{eq:evolutionsolutionLCplus}
{\cal U}^{{\rm LC+}}(t,t_0) = 
{\cal N}^{{\rm LC+}}(t,t_0)
+ \int_{t_0}^t\!d\tau\ 
{\cal U}^{{\rm LC+}}(t,\tau)\,
{\cal H}_I^{{\rm LC+}}(\tau)\,
{\cal N}^{{\rm LC+}}(\tau,t_0) 
\;\;.
\end{equation}
Here ${\cal N}^{{\rm LC+}}(t_{2},t_1)$ is the no-splitting operator,
\begin{equation}
{\cal N}^{{\rm LC+}}(t_2,t_1) = \exp\left[
-\int_{t_1}^{t_2} d\tau\ {\cal V}^{{\rm LC+}}(\tau)\right]
\;\;.
\end{equation}
That is, ${\cal N}^{{\rm LC+}}(t_2,t_1)$ is the Sudakov factor that represents the probability that no parton splits between shower time $t_1$ and shower time $t_2$. The essential point of the LC+ approximation is that the operator ${\cal V}^{{\rm LC+}}(\tau)$ is diagonal in the standard color basis that we use, so that it is practical to calculate its exponential. In contrast, the operator ${\cal V}(\tau)$ is not diagonal in color, so that it is not practical to evaluate its exponential.

Now, what if we want shower evolution with full color? Then we need
\begin{equation}
\label{eq:evolutionU2}
\frac{d}{dt}\,{\cal U}(t,t_0)
= [{\cal H}_I^{{\rm LC+}}(t) - {\cal V}^{{\rm LC+}}(t) + \Delta {\cal H}_I(t) - \Delta{\cal V}(t)]\,
{\cal U}(t,t_0)
\;\;,
\end{equation}
where
\begin{equation}
\begin{split}
\Delta {\cal H}_I(t) ={}& {\cal H}_I(t) - {\cal H}_I^{{\rm LC+}}(t)
\;\;,
\\
\Delta {\cal V}(t) ={}& {\cal V}(t) - {\cal V}^{{\rm LC+}}(t)
\;\;.
\end{split}
\end{equation}
This evolution equation gives
\begin{equation}
\label{eq:evolutionsolutionfull}
{\cal U}(t,t_0) = 
{\cal U}^{{\rm LC+}}(t,t_0)
+ \int_{t_0}^t\!d\tau\ 
{\cal U}(t,\tau)\,
\left[
\Delta {\cal H}_I(\tau)
-\Delta {\cal V}(\tau)
\right]
{\cal U}^{{\rm LC+}}(\tau,t_0) 
\;\;,
\end{equation}
which can be solved iteratively:
\begin{equation}
\begin{split}
\label{eq:softexpansion}
{\cal U}(t_\Lf,t_0) ={}& {\cal U}^{{\rm LC+}}(t_\Lf,t_0)
\\&+
 \int_{t_0}^{t_\Lf}\! d\tau\ 
{\cal U}^{{\rm LC+}}(t,\tau_1)\,
\big[\Delta {\cal H}_I(\tau_1)
- \Delta {\cal V}(\tau_1)\big]\,
{\cal U}^{{\rm LC+}}(\tau_1,t_0)
\\&+
 \int_{t_0}^{t_\Lf}\! d\tau_2
 \int_{t_0}^{\tau_2}\! d\tau_1\
{\cal U}^{{\rm LC+}}(t_\Lf,\tau_2)\,
\big[\Delta {\cal H}_I(\tau_2)
- \Delta {\cal V}(\tau_2)\big]\,
{\cal U}^{{\rm LC+}}(\tau_2,\tau_1)
\\&\qquad\times
\big[\Delta {\cal H}_I(\tau_1)
- \Delta {\cal V}(\tau_1)\big]\,
{\cal U}^{{\rm LC+}}(\tau_1,t_0)
\\ &+\cdots
\;\;.
\end{split}
\end{equation}
One can have any (small) maximum number of insertions of $\big[\Delta {\cal H}_I(\tau)- \Delta {\cal V}(\tau)\big]$. In between the insertions, we have LC+ evolution. The operator $\big[\Delta {\cal H}_I(\tau)- \Delta {\cal V}(\tau)\big]$ can change the color state of the system, but we can still proceed because ${\cal U}^{{\rm LC+}}(\tau_{i+1},\tau_i)$ can act on any color state.

We will argue in section \ref{sec:gapfraction} that this sort of systematic improvement is needed for some problems. We have, however, not implemented this feature in the current version of \textsc{Deductor}.

\section{How much color suppression is typical?}
\label{sec:howmuchI}

\begin{figure}
\centerline{\includegraphics[width = 10 cm]{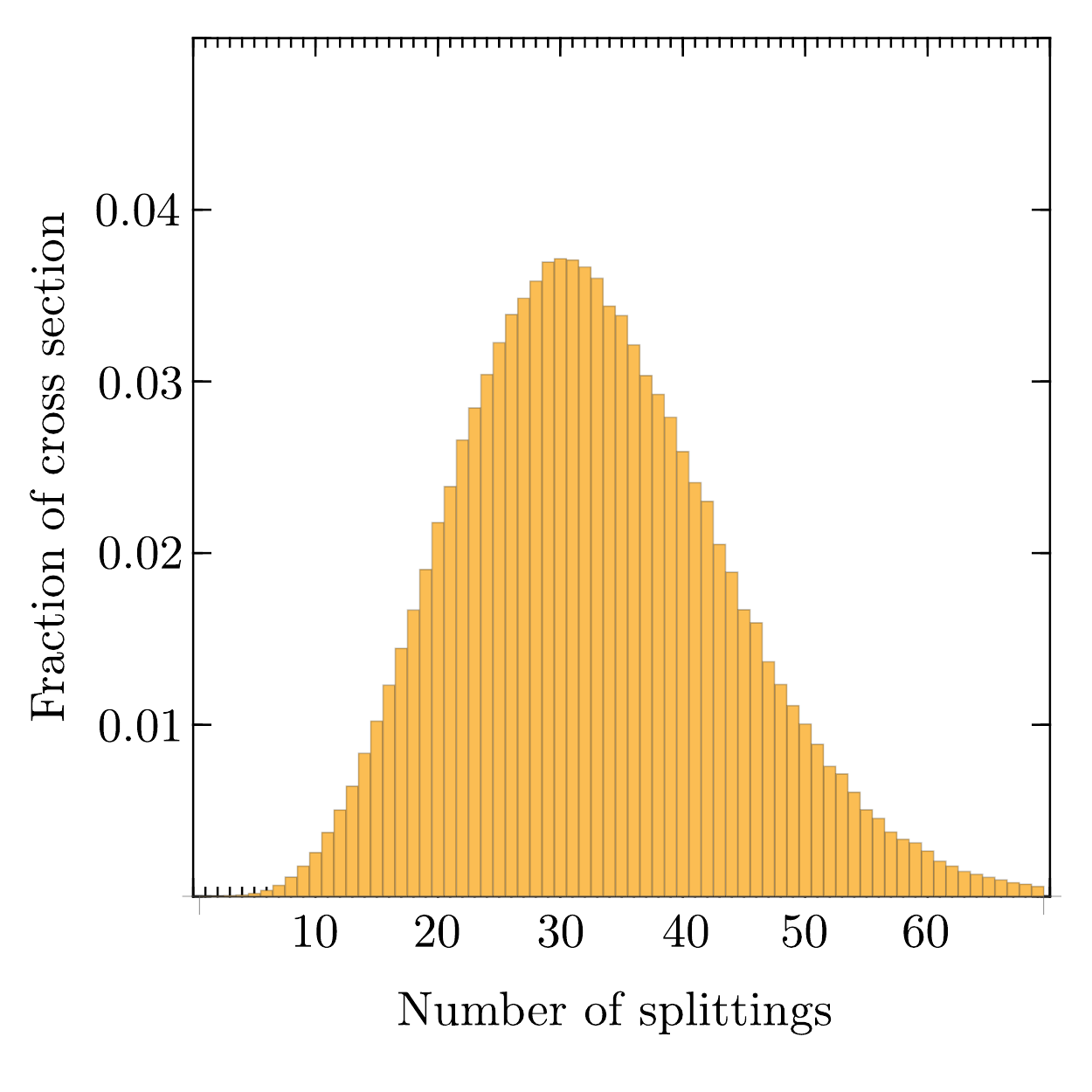}}

\caption{Probability of generating $N$ splittings versus $N$ in events that have a jet with $p_\LT > 500 \GeV$ in the rapidity range $-2 < y < 2$. The splitting cutoff is $k_\LT^{\rm min} = 1 \GeV$.}
\label{fig:Nsplittings}
\end{figure}

Is it rare to generate a state with non-zero color suppression index $I$? To find out, we use \textsc{Deductor} to generate the cross section to make jets in proton-proton collisions at $\sqrt s = 14 \TeV$. We select events that have a jet with $p_\LT > 500 \GeV$ in the rapidity range $-2 < y < 2$. The parton shower generates splittings that are softer and softer, as measured by the virtuality parameter $\Lambda^2$ defined in ref.~\cite{ShowerTime}. We stop the shower by not allowing any splittings with a splitting transverse momentum smaller than $k_\LT^{\rm min} = 1 \GeV$. This allows for quite a number of splittings: most of the cross section comes from events with twenty to fifty splittings. We show the distribution of the number of splittings in figure \ref{fig:Nsplittings}.

The hard process here is parton-parton scattering.\footnote{We have checked our calculation of the color density matrix for parton-parton scattering against the results of Kuijf \cite{Kuijf}.} We keep all of the contributions to the color-ordered amplitudes for parton-parton scattering. This gives an initial color density matrix of the form (\ref{eq:rhodef}) with $m = 2$ for the two final state partons. We calculate the color suppression index $I_2$ just after the hard scattering according to eq.~(\ref{eq:colorindex}) with $p_\LE = 0$.  We find that, including the cases with $\{c'\}_2 \ne \{c\}_2$, the color suppression index can be $I_2 = 0$, 1, or 2. Now the shower starts. With each splitting, $I$ has a chance to increase, so that $I_m \ge I_2$. We set a limit $I_m - I_2 \le (\Delta I)_{\rm max}$ with $(\Delta I)_{\rm max} = 4$. If $I_m - I_2$ reaches $I_{\rm max}$, we switch off the LC+ approximation and start using the extended version of the LC approximation that allows for $\{c'\}_m \ne \{c\}_m$.

\begin{figure}
\centerline{\includegraphics[width = 10 cm]{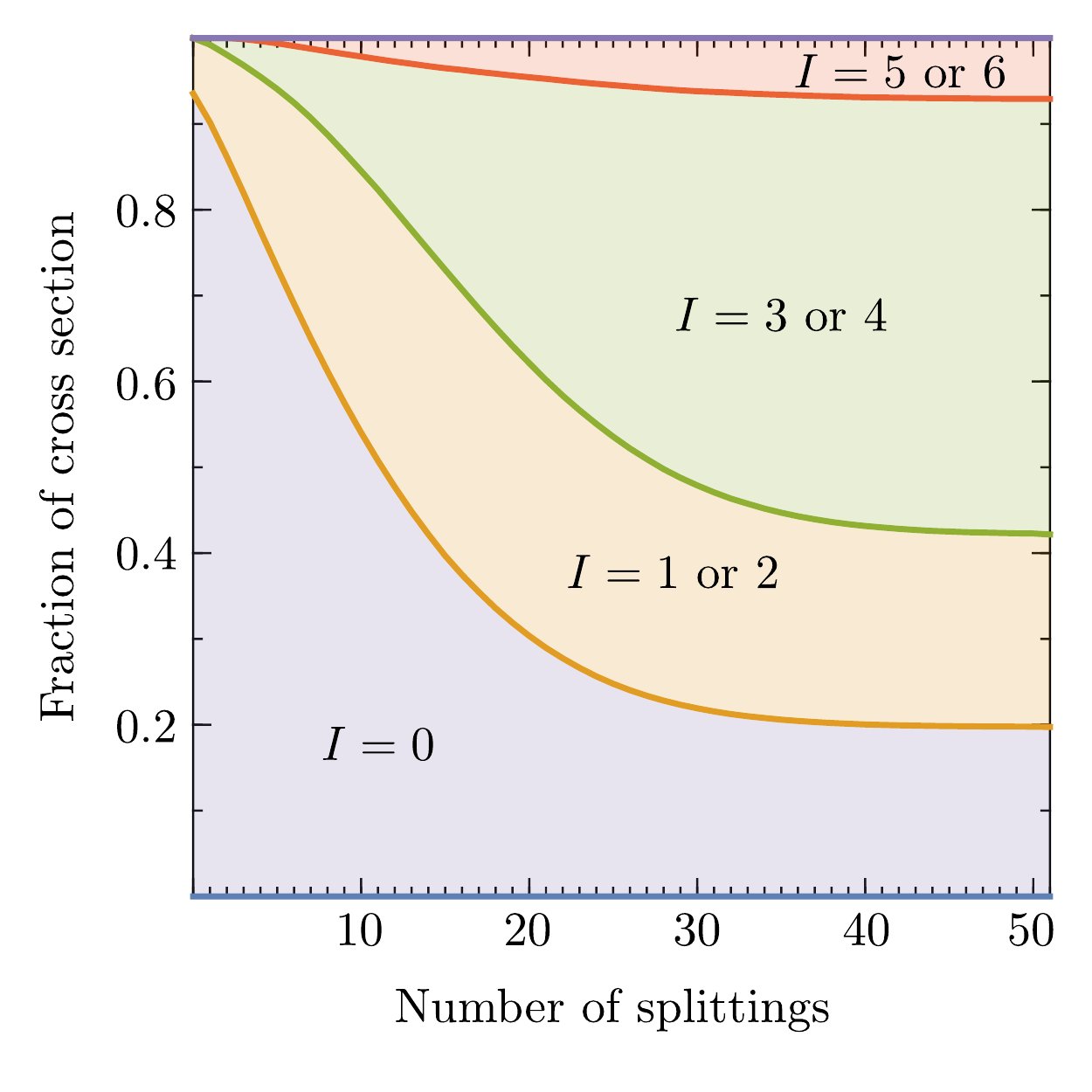}}

\caption{Fraction of events with given values, $I$, of the color suppression index as a function of the number of splittings. We set a limit of 4 on the change of $I$ after the hard scattering.
}
\label{fig:colorgrowth}
\end{figure}

We can answer the question of how rare it is to generate a non-zero color suppression index by plotting the fraction of the jet cross section coming from events that have a given value of $I_{N+2}$ after $N$ splittings. We group $I_{N+2} = 1$ with $I_{N+2} = 2$, $I_{N+2} = 3$ with $I_{N+2} = 4$, and $I_{N+2} = 5$ with $I_{N+2} = 6$ because odd values of $I$ (which are generated at the hard scattering) are rather rare. We plot the fraction of events with given values of $I_{N+2}$ for $N$ up to 50. For events with $S$ splittings with $S < 50$, we define $I_{N+2} = I_{S + 2}$ for $N > S$. We show a plot of the fraction of events with given values of $I_{N+2}$ after $N$ splittings in figure \ref{fig:colorgrowth}.

We can draw two lessons from figure \ref{fig:colorgrowth}. First, a little less than 10\% of the cross section is associated with events that start with $I > 0$ right at the hard scattering. This is about what one would expect, since $1/N_\Lc^2 \approx 0.1$. Second, it is rather rare for the partons to stay in an $I = 0$ state after many splittings. Only about 20\% of the cross section comes from parton configurations that have $I = 0$ at the end of the shower. In fact, more than half of the cross section has $I > 2$ at the end of the shower. We would see more cross section associated with $I > 4$ at the end of the shower if we had not imposed $I_m - I_2 \le 4$ on the shower.

\section{Jet cross section}
\label{sec:jetcrosssection}

We have seen that most of the cross section to make jets in proton-proton collisions at $\sqrt s = 14 \TeV$ is associated with partonic states at the end of the parton shower that have color suppression index $I > 0$. Does this matter for the jet production cross section?

One would expect that it does not matter, since the jet production cross section is so inclusive. That is, by design, the cross section to make a jet with a certain $p_\LT$ in a rapidity range $-2 < y < 2$ is very much infrared safe. The cross section is mostly determined by the cross section for parton-parton scattering (calculated at the Born level in this case). 

\begin{figure}
\centerline{\includegraphics[width = 10 cm]{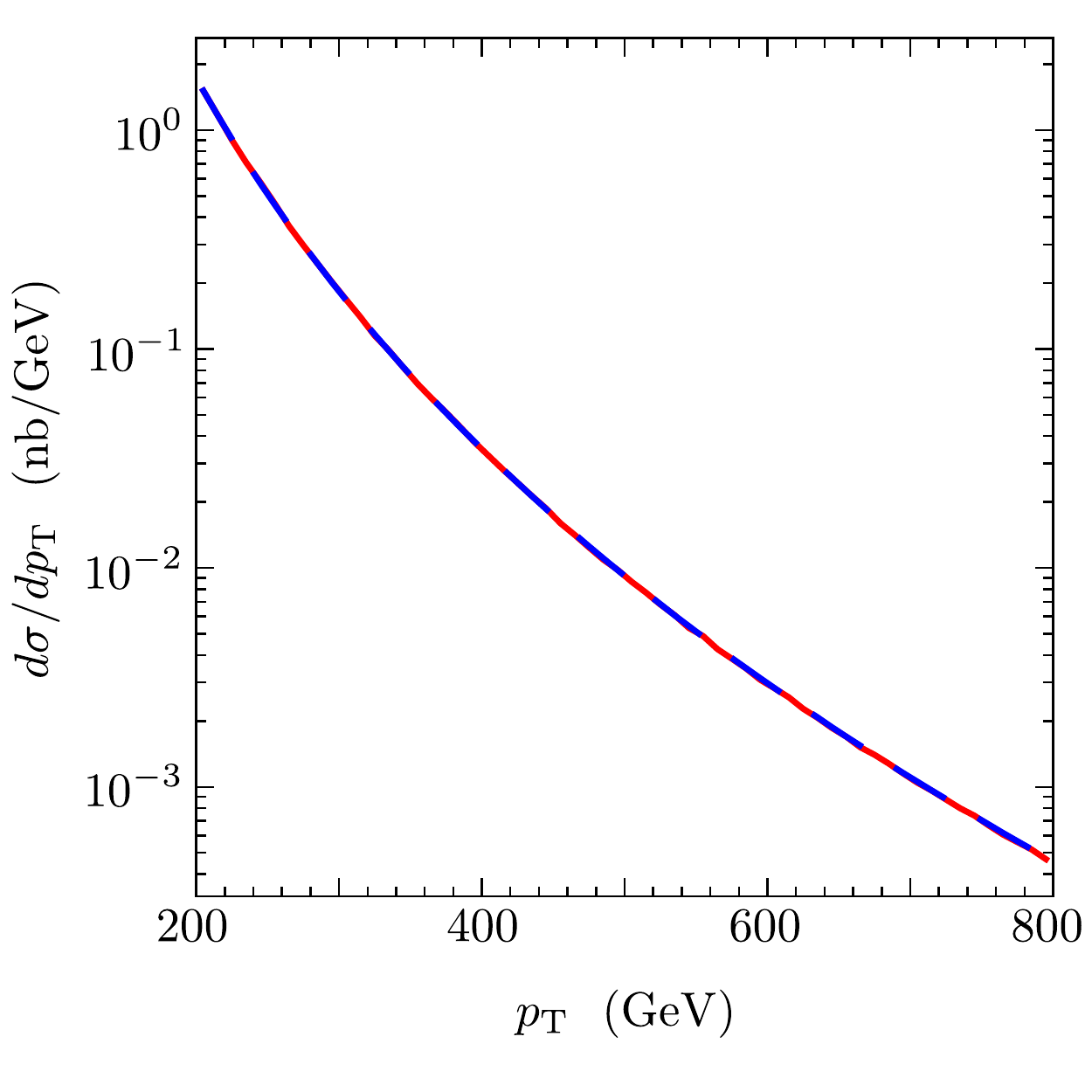}}

\caption{One jet inclusive cross section $d\sigma/dp_\LT$ for $|y|<2$ using the $k_\LT$ algorithm with $R = 0.4$. The red solid line shows the cross section calculated with the LC+ approximation. The blue dashed line shows the cross section calculated with the LC approximation for the parton shower. The two curves are indistinguishable.
}
\label{fig:onejet}
\end{figure}

Parton shower evolution can add some $p_\LT$ to the jet from partons radiated from the initial-state partons. Shower evolution can also subtract some $p_\LT$ from the jet when the scattered parton that initiates the jet radiates a gluon at such a wide angle that the gluon is not included in the jet. This initial-state and final-state radiation, along with the interference between initial-state and final-state radiation, is influenced by the color configuration of the radiating partons. Thus the development of the color configuration can influence the ultimate jet cross section. Since the color state develops differently in the LC+ approximation than it does in the LC approximation, the jet cross section could be affected. However, since the jet cross section is designed to be insensitive to all of these infrared or long-time effects, one does not expect to see much difference between the jet cross section calculated in the LC+ approximation and the jet cross section calculated in the LC approximation.

We can test the hypothesis that the LC approximation gives a jet cross section close to that of the LC+ approximation. We use \textsc{Deductor} to calculate the one jet inclusive cross section $d\sigma/dp_T$ to produce a jet with transverse momentum $p_\LT$ in the rapidity range $-2 < y < 2$. With the help of \textsc{FastJet} \cite{FastJet}, we use the $k_\LT$ jet algorithm \cite{KTCatani,KTEllis} with $R = 0.4$. The results are shown in figure \ref{fig:onejet}. The solid line shows the cross section calculated with the LC+ approximation with $(\Delta I)_{\rm max} = 4$. The dashed line shows the result obtained with the LC approximation.\footnote{In the hard scattering, density matrix contributions $\ket{\{c\}_2}\bra{\{c'\}_2}$ with ${\{c\}_2} \ne {\{c'\}_2}$ are eliminated and the corresponding cross section is redistributed to density matrix contributions $\ket{\{c\}_2}\bra{\{c'\}_2}$ with ${\{c\}_2} = {\{c'\}_2}$. Then $(\Delta I)_{\rm max}$ is set to zero in the shower so that the color suppression index stays equal to zero.} The dashed line and the solid line are indistinguishable on the graph. We would show a graph of the ratio of the two cross sections, but we find that the difference is less than 2\%, which is the statistical accuracy of the calculation.

We conclude that the use of the LC approximation for the shower instead of the LC+ approximation makes a negligible difference for the one jet inclusive cross section, even though most of the cross section comes from states that have color suppression index $I > 0$ by the end of the shower. This is not a surprising conclusion, but it is not a conclusion that one could be sure of without doing the calculation.

\section{Number of partons in a jet}
\label{sec:NinJet}

In section \ref{sec:jetcrosssection} we looked at an observable that is very insensitive to soft parton splittings: the one jet inclusive cross section. Now, we look inside these jets at a quantity that is {\em sensitive} to soft parton splittings: the distribution of the number of partons in a jet. Evidently, the number of partons in a jet is not a physical observable, but it is calculable with a fixed cut on the smallest transverse momentum allowed in a parton splitting. It is a stand-in for the number of hadrons in a jet, which is physically observable. The distribution of the number of partons in a jet is of interest here because it is an infrared sensitive quantity.

We analyze a sample of jets with $p_\LT > 200 \GeV$ and $|y| < 2$. We examine the distribution $\rho_n(n)$ of the number $n$ of partons in a jet in this sample for events simulated by \textsc{Deductor} using the LC+ approximation and using the LC approximation. The distribution is normalized to $\sum_n \rho_n(n) = 1$. In each case, we stop the shower by not allowing any splittings with a splitting transverse momentum smaller than $k_\LT^{\rm min} = 1 \GeV$. 

\begin{figure}
\centerline{\includegraphics[width = 8 cm]{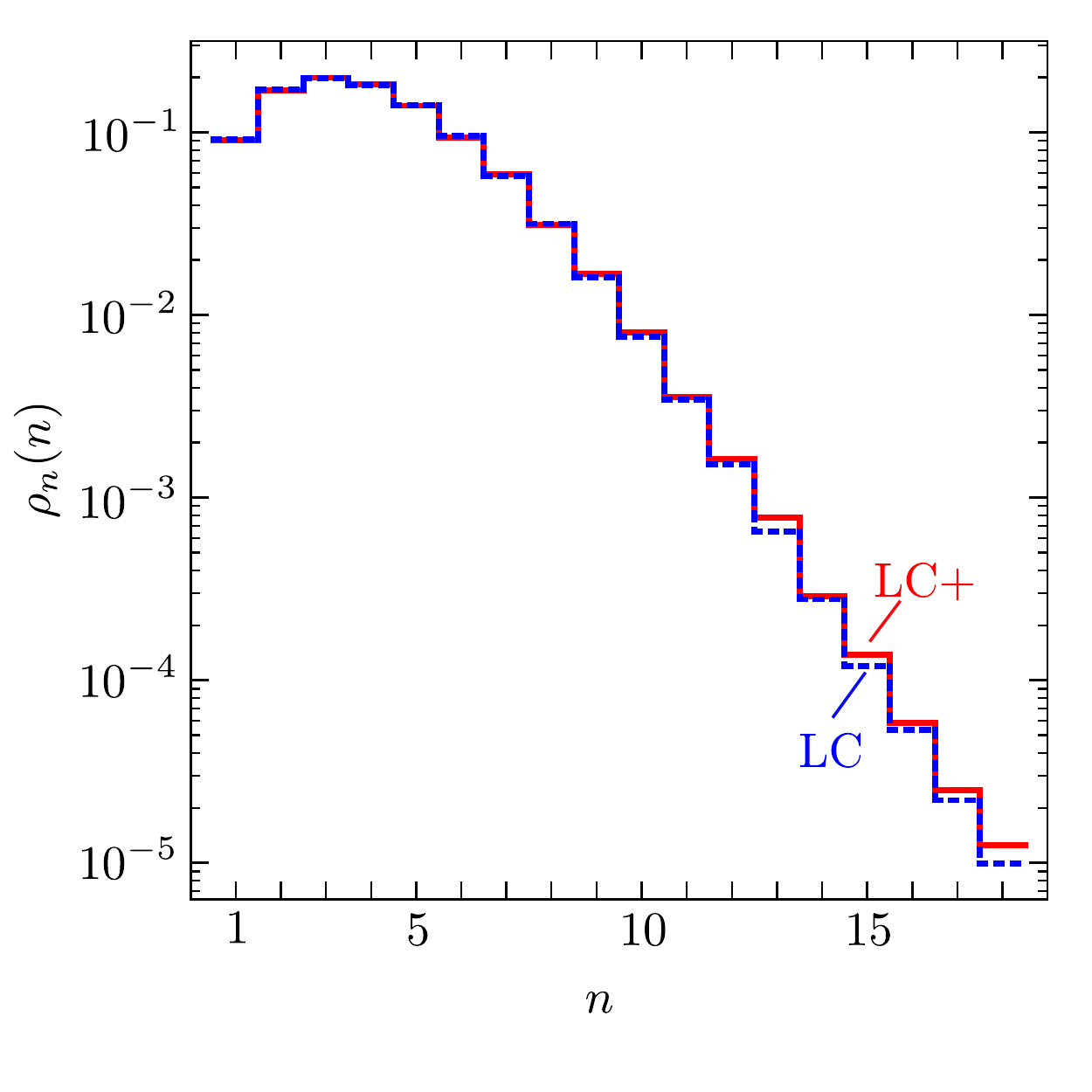}}

\caption{The distribution $\rho_n(n)$ for the number $n$ of partons in a jet with $p_\LT > 200 \GeV$ and $|y| < 2$. We compare the distribution calculated with the LC+ approximation (solid red histogram) with the  with the same distribution calculated with the LC approximation (dashed blue histogram). The results are very close for $n < 10$. The jets are constructed using the $k_\LT$ algorithm with $R = 0.4$.
}
\label{fig:Ninjet}
\end{figure}

The result is displayed in figure \ref{fig:Ninjet}. We see that for $n < 10$, $\rho_n(n)$ is quite insensitive to whether we use the LC or LC+ approximation.  The behavior of $\rho_n(n)$ for large $n$ is of special interest because when there are many partons, most of them must have small momenta, so that for large $n$ we probe the softest emissions. The probability that there are more than 10 partons in a jet is small. We notice that for $n > 10$, $\rho_n(n)$ calculated using the LC+ approximation is about 10\% larger than $\rho_n(n)$ calculated using the LC approximation. The fractional difference is of order $1/N_\Lc^2$, in accordance with the simplest expectation.

\section{Gap fraction}
\label{sec:gapfraction}

We now turn to an observable for which we believe that the color treatment matters: the probability that there are no jets in the rapidity range between two jets that are widely separated in rapidity.

In proton-proton collisions at $\sqrt s = 14 \TeV$, we look for jets using the $k_\LT$ algorithm with $R = 0.4$. We select events for which there is at least one jet with transverse momentum $p_\LT > 200 \GeV$ and rapidity $y > 2$ and at least one jet with transverse momentum $p_\LT > 200 \GeV$ and rapidity $y < -2$. For a given transverse momentum $p_{\LT}^{\rm cut}$, we can ask for the fraction $f(p_{\LT}^{\rm cut})$ of selected events in which no jet that has transverse momentum greater than $p_{\LT}^{\rm cut}$ appears in the rapidity region $-2 < y < 2$. 

There has been considerable work on the theory of this sort of observable \cite{EarlyGap1, EarlyGap2, Manchester2005, NonGlobal1, NonGlobal2, NonGlobal3, NonGlobal4, Manchester2009, SuperLeading1, SuperLeading2, Manchester2011}. The treatment of the problem in an angle ordered parton shower is analyzed in ref.~\cite{SchofieldSeymour}. 

\begin{figure}
\centerline{\includegraphics[width = 8 cm]{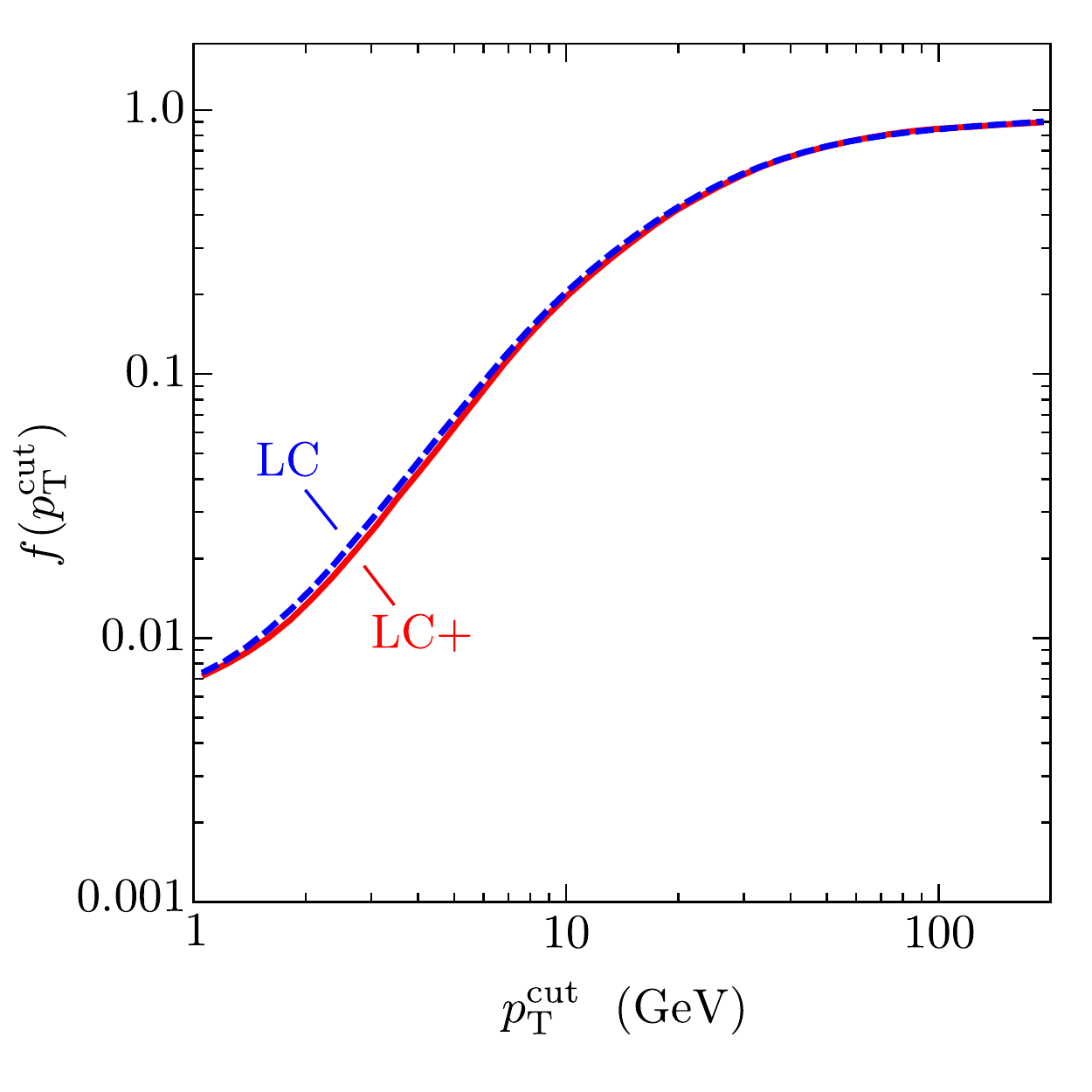}}

\caption{For events with at least one jet with transverse momentum $p_\LT > 200 \GeV$ and rapidity $y > 2$ and at least one jet with transverse momentum $p_\LT > 200 \GeV$ and rapidity $y < -2$, we plot the fraction $f(p_{\LT}^{\rm cut})$ of events that have no jets with transverse momentum greater than $p_{\LT}^{\rm cut}$ in the rapidity range $-2 < y < 2$. The blue dashed curve shows the result obtained with the LC approximation and the red solid curve shows the result obtained with the LC+ approximation.
}
\label{fig:gap1}
\end{figure}

In figure \ref{fig:gap1}, we plot $f(p_{\LT}^{\rm cut})$, versus $p_{\LT}^{\rm cut}$. For large values of $p_{\LT}^{\rm cut}$, the probability to produce an extra jet is suppressed by a factor $\as$ and by the need to have parton distribution functions with larger momentum fraction arguments, so that we expect $f(p_{\LT}^{\rm cut})$ to be close to 1. However, it is typically rather easy to produce low transverse momentum jets, so for small $p_{\LT}^{\rm cut}$, we expect $f(p_{\LT}^{\rm cut})$ to be small. This is what we see in figure \ref{fig:gap1}. We have calculated $f(p_{\LT}^{\rm cut})$ both in the LC approximation and in the LC+ approximation. We see that there is about a 10\% difference between the two results at small values of $p_{\LT}^{\rm cut}$. This level of difference is similar to what we saw for the distribution of the number of partons in a jet in section \ref{sec:NinJet}. 

This close agreement between the LC approximation and the LC+ approximation seems to suggest that color flow issues are not important for the gap fraction. However, we believe that such a view would be mistaken and that both the LC approximation and the LC+ approximation are inadequate for this problem. We can understand some of the issues without going into detail as follows. 

\begin{figure}
\centerline{\includegraphics[width = 6 cm]{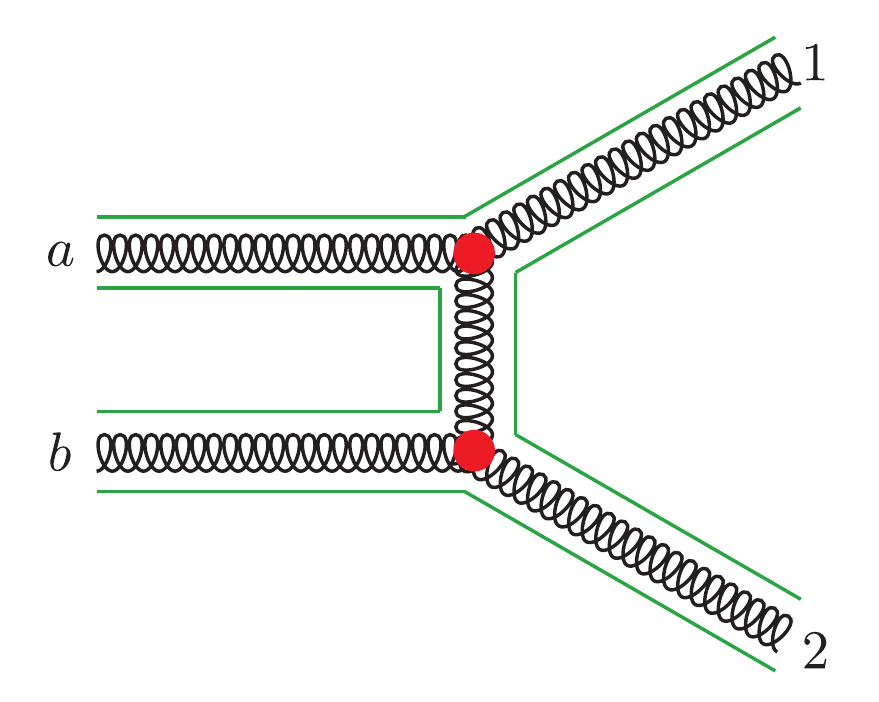}}

\caption{Gluon-gluon scattering via gluon exchange}
\label{fig:ggscatt}
\end{figure}

The rate of decrease of $f$ as $p_{\LT}^{\rm cut}$ decreases is controlled by the color flow in the event. To see this from the point of view of a dipole based parton shower, consider gluon-gluon scattering, as depicted in figure \ref{fig:ggscatt}. Partons ``a'' and ``b'' scatter to produce final-state partons 1 and 2. Parton ``a'' has infinite positive rapidity, while parton ``b'' has infinite negative rapidity. In a leading color picture, partons ``a'' and ``b'' are color connected, as are ``a'' and 1, ``b'' and 2, and 1 and 2, as indicated in figure \ref{fig:ggscatt}. There is another color configuration in which ``a'' is connected to 2 and ``b'' is connected to 1.

These parton pairs form color dipoles. In a parton shower, the a-1 pair produces soft radiation in the angular region between $\vec p_\La$ and $\vec p_1$. That is, this radiation has large positive rapidity. Similarly, the b-2 dipole produces soft radiation with large negative rapidity. The a-b dipole produces soft radiation with any rapidity between a large positive value and a large negative value. Similarly, the 1-2 dipole produces soft radiation with any rapidity between the large positive rapidity of parton 1 and a large negative rapidity of parton 2. Thus it is very likely that the a-b, 1-2, a-2, and b-1 dipoles will produce radiation that fills in the gap.

\begin{figure}
\centerline{\includegraphics[width = 6 cm]{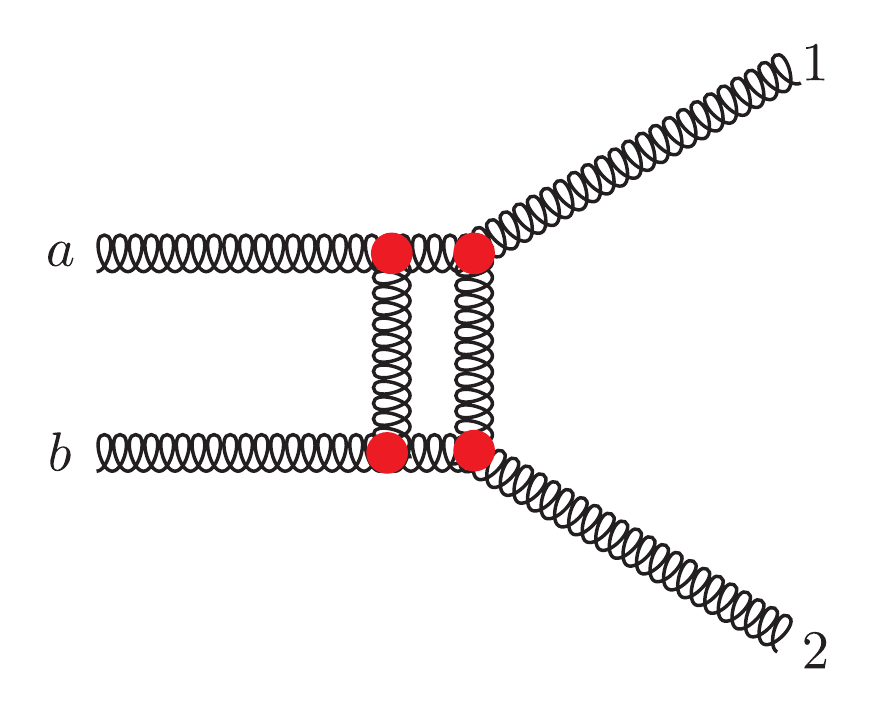}}

\caption{A virtual correction to gluon-gluon scattering}
\label{fig:ggscattVirt}
\end{figure}

Now consider what happens in a parton shower based on the LC+ approximation with perturbative corrections as in eq.~(\ref{eq:softexpansion}). There can be a $\Delta {\cal V}(\tau)$ contribution that represents a virtual gluon exchange between the two incoming partons, including imaginary contributions proportional to $\mi \pi$. This gives a diagram in which a hard gluon is exchanged, then a soft gluon, as depicted in fig.~\ref{fig:ggscattVirt}. The exchange of two gluons is equivalent in color space to the exchange of an object with color $\bm 27$, color $\bm {10}$, color $\overline{\bm {10}}$, color $\bm 8$ (in two ways), or color $\bm 1$. With color singlet exchange, the two forward moving partons, ``a'' and 1, are completely disconnected in color space from the two backward moving partons, ``b'' and 2. If we get this color configuration in both the color ket state and the color bra state, we get forward radiation and backward radiation, but only a small amount of radiation into the gap region. Contributions like this can change the gap fraction. 

This same argument applies to virtual gluon exchange between partons 1 and 2. It also applies to virtual gluon exchange between ``a'' and 2 or between ``b'' and 1. It also applies to $q$-g scattering, $q$-$q$ scattering, and $q$-$\bar q$ scattering.

It is clear that this effect is not included in the LC+ shower with no factors of $\Delta {\cal V}(\tau)$. The existing literature \cite{EarlyGap1, EarlyGap2, Manchester2005, NonGlobal1, NonGlobal2, NonGlobal3, NonGlobal4, Manchester2009, SuperLeading1, SuperLeading2, Manchester2011, SchofieldSeymour} suggests, however, that the effect is important. To include it, we need to use the perturbative expansion of eq.~(\ref{eq:softexpansion}). This expansion is not yet implemented in \textsc{Deductor}.

\section{Conclusions and outlook}
\label{sec:conclusions}

We have examined the effect of color flow on the results from a parton shower generator based on color dipoles, \textsc{Deductor}. On one hand, we used the leading color, LC, approximation that is standard for parton shower generators. We compared LC results to results from the LC+ approximation that is available in \textsc{Deductor}. We used the LC+ approximation up to terms suppressed by more than four powers of $1/N_\Lc$. The LC+ approximation has the advantage that it is exact in color for collinear emissions. It is, however, only approximate for wide angle soft gluon emissions. 

By comparing LC results to LC+ results, we tested numerically how good the LC approximation is. We found that the LC approximation is quite good for the one jet inclusive cross section and for the distribution of the number of partons in a jet. There is an important caution needed here: one would really need to test the LC approximation against a full color treatment to be sure that the LC approximation is accurate for any given observable. However, testing the LC approximation against the LC+ approximation certainly gives the LC approximation a chance to fail and it does not. We have tried some other observables and have found similar results.

In section \ref{sec:howmuchI}, we asked how much color suppression is typical. We found that color evolution beyond the leading color approximation is not rare. It is the norm. However, we find that simple observables are not much affected by deviations from leading color configurations.

In one case, we found that the LC approximation agrees well with the LC+ approximation, but we believe that neither approximation is adequate: the gap-between-jets cross section investigated in section \ref{sec:gapfraction}. Our belief here is based on the analyses in the substantial current literature \cite{EarlyGap1, EarlyGap2, Manchester2005, NonGlobal1, NonGlobal2, NonGlobal3, NonGlobal4, Manchester2009, SuperLeading1, SuperLeading2, Manchester2011, SchofieldSeymour}: the color changing operators that need to be present are in fact not present in the LC+ approximation.

We believe that a parton shower approach to the gap-between-jets problem holds some promise because a parton shower approach can sometimes provide numerical results to complicated problems that are too unwieldy to solve analytically. Certainly, a parton shower can incorporate the effects of momentum conservation in the emission of soft gluons. These effects are not easy to incorporate in an analytical approach and they can be important \cite{SchofieldSeymour}. In addition, a parton shower automatically includes potential radiation into the gap from a gluon that was radiated out of the gap. What is needed is a better treatment of color within the parton shower. In ref.~\cite{NScolor}, we described how one can insert perturbatively the evolution operators that represent the difference between exact color and the LC+ approximation. See eq.~(\ref{eq:softexpansion}). It is perhaps not necessary to have a large number of insertions in order to have a reasonably good approximation: one virtual gluon exchange in the ket state and one in the bra state may be enough, as suggested by the numerical investigations in ref.~\cite{Platzer}. We have not, however, implemented such a program, so we do not know how well it will work.

Our tests do not include hadronization. We anticipate in the future linking \textsc{Deductor} to the string hadronization model \cite{LundString1, LundString2} of \textsc{Pythia}. For this purpose, we need to specify the probability for a given color density matrix element $\ket{\{c\}_m}\bra{\{c'\}_m}$ from eq.~(\ref{eq:rhodef}) to correspond to a given classical string state. A method for doing this is given in section 8 of ref.~\cite{NScolor}. We expect that the string configurations produced with the LC+ approximation may have more kinks than those produced with the LC approximation. Presumably this will necessitate retuning the parameters of the hadronization model compared to the \textsc{Pythia} default tune. It is difficult to predict what effects may remain after retuning.

Finally, we note that a shower based on the LC+ approximation can start with any basis element $\ket{\{c\}_m}\bra{\{c'\}_m}$ of the color density matrix, including one with $\{c'\}_m \ne \{c\}_m$. This is not the case with the LC approximation as usually formulated. This is advantageous for matching to a perturbative NLO calculation since one can use the full NLO color-ordered amplitudes. There is a matching program \cite{Czakon} that matches to \textsc{Deductor} in an MC@NLO style. As presently implemented, this calculation reduces the NLO density matrix to $\{c'\}_m = \{c\}_m$, but we expect that this can be generalized to $\{c'\}_m \ne \{c\}_m$.

\acknowledgments{ 
This work was supported in part by the United States Department of Energy and by the Helmholtz Alliance ``Physics at the Terascale." We thank Jeff Forshaw and Mrinal Dasgupta for helpful conversations about rapidity gap physics.}

\appendix

\section{Using the leading color approximation}
\label{sec:LC}

\textsc{Deductor} uses the LC+ approximation for color, as explained in some detail in ref.~\cite{NScolor}. However, we may want to use just an LC shower. In particular, it is useful to turn the LC+ shower off if the amount that the color suppression index has increased during the shower (after the hard interaction) reaches a preset value, $(\Delta I)_{\rm max}$. After that, we can proceed with an LC shower. The leading color approximation is commonly used in parton shower algorithms, so it would hardly need an explanation. However, we need to run an LC shower starting with a color density matrix 
\begin{equation}
\label{eq:rhodefbis}
\rho(\{p,f\}_m,t) 
= \sum_{\{c\}_m ,\{c'\}_m}\rho(\{p,f,c',c\}_m,t)\,\ket{\{c\}_m}\bra{\{c'\}_m}
\;,
\end{equation}
where possibly $\{c'\}_m \ne \{c\}_m$. This is straightforward, but it is not commonly done, so we explain it in this appendix. In order to keep the appendix brief, we assume that the material in refs.~\cite{NSI} and \cite{NScolor} is known.

The method for switching from LC+ to LC that we explain here and that is implemented in \textsc{Deductor} is different from the method outlined in appendix C of ref.~\cite{NScolor}.

The LC approximation is based on replacing the color group $SU(N_\Lc)$ by $U(N_\Lc)$. There are $N_\Lc^2$ gluons, labeled with $q = \{1,2,\dots, N_\Lc^2\}$. The generator matrices $t^1,\dots,t^{N_\Lc^2-1}$ in the fundamental representation are the usual ones. The new generator is
\begin{equation}
\label{eq:t9def}
t_{i i'}^{N_\Lc^2} = \frac{1}{\sqrt{2N_\Lc}}\ \delta_{i i'}
\;.
\end{equation}
This gives the identity
\begin{equation}
\label{eq:U3Fierz}
\sum_{a = 1}^{N_\Lc^2} t_{i i'}^a t_{j j'}^a = \frac{1}{2}\ \delta_{i j'} \delta_{j i'}
\;.
\end{equation}
That is, aside from the factor 1/2, we can think of each gluon as carrying one color $N_\Lc$ line and one color $\bar N_\Lc$ line. The factor $T_\LR = 1/2$ comes from our  conventional normalization of the generator matrices,
\begin{equation}
{\rm Tr}[t^a t^b] = \frac{1}{2}\,\delta_{ab}
\;.
\end{equation}
With the definition Eq.~(\ref{eq:t9def}), this normalization remains true for all $a$ and $b$.

We define the structure constants $f_{abc}$ that give the couplings of the gluons to each other by
\begin{equation}
\label{eq:f_abc}
[t^a,t^b] = \mi f_{abc} t^c
\;.
\end{equation}
With this definition, we see that the $U(1)$ gluon that we have added, with index $N_\Lc^2$, does not couple to the other gluons. That is, $f_{abc} = 0$ if any of $a$, $b$, or $c$ is $N_\Lc^2$.

Our analysis makes use of the standard color basis defined by two sorts of vectors, as in ref.~\cite{NSI}. First, there are open string vectors
\begin{equation}
\label{eq:openstring}
\Psi(S)^{\{ a \}} 
= N(S)^{-1/2} [t^{a_2} t^{a_3}\cdots t^{a_{n-1}}]_{a_1,a_n}
\;\;.
\end{equation}
Here $a_1$ is a quark color index, $a_n$ is an antiquark color index, and the other $a_i$ are gluon color indices. The $t^a$ are $U(N_\Lc)$ color matrices in the fundamental representation. Also, $N(S) = N_\Lc (N_\Lc/2)^{n-2}$ is a normalization factor such that $\brax{\Psi}\ket{\Psi} = 1$. Second, there are closed string vectors
\begin{equation}
\label{eq:closedstring}
\Psi(0)^{\{ a\}} 
= N(S)^{-1/2}\, {\rm Tr}[t^{a_1} t^{a_2}\cdots t^{a_{n}}]
\;\;.
\end{equation}
Here all of the $a_i$ are gluon color indices. Again, $N(S) = (N_\Lc/2)^{n}$ is a normalization factor such that $\brax{\Psi}\ket{\Psi} = 1$. (With $SU(N_{\Lc})$, $\brax{\Psi}\ket{\Psi}$ is slightly different from 1.) The most general color basis vector, which we denote by $\ket{\{c\}_m}$, is a product of these two kinds of units.

The color basis states are normalized to $\brax{\{c\}_m}\ket{\{c\}_m} = 1$. They are not, however, generally orthogonal. However, when $\{c\}_m$ and $\{c'\}_m$ are different, one finds that $\brax{\{c'\}_m}\ket{\{c\}_m} = {\cal O}(1/N_\Lc^2) \ {\rm or}\ {\cal O}(1/N_\Lc)$. That is, the basis vectors are orthogonal only in the $N_\Lc \to \infty$ limit.

\begin{figure}
\centerline{\includegraphics[width = 13.6 cm]{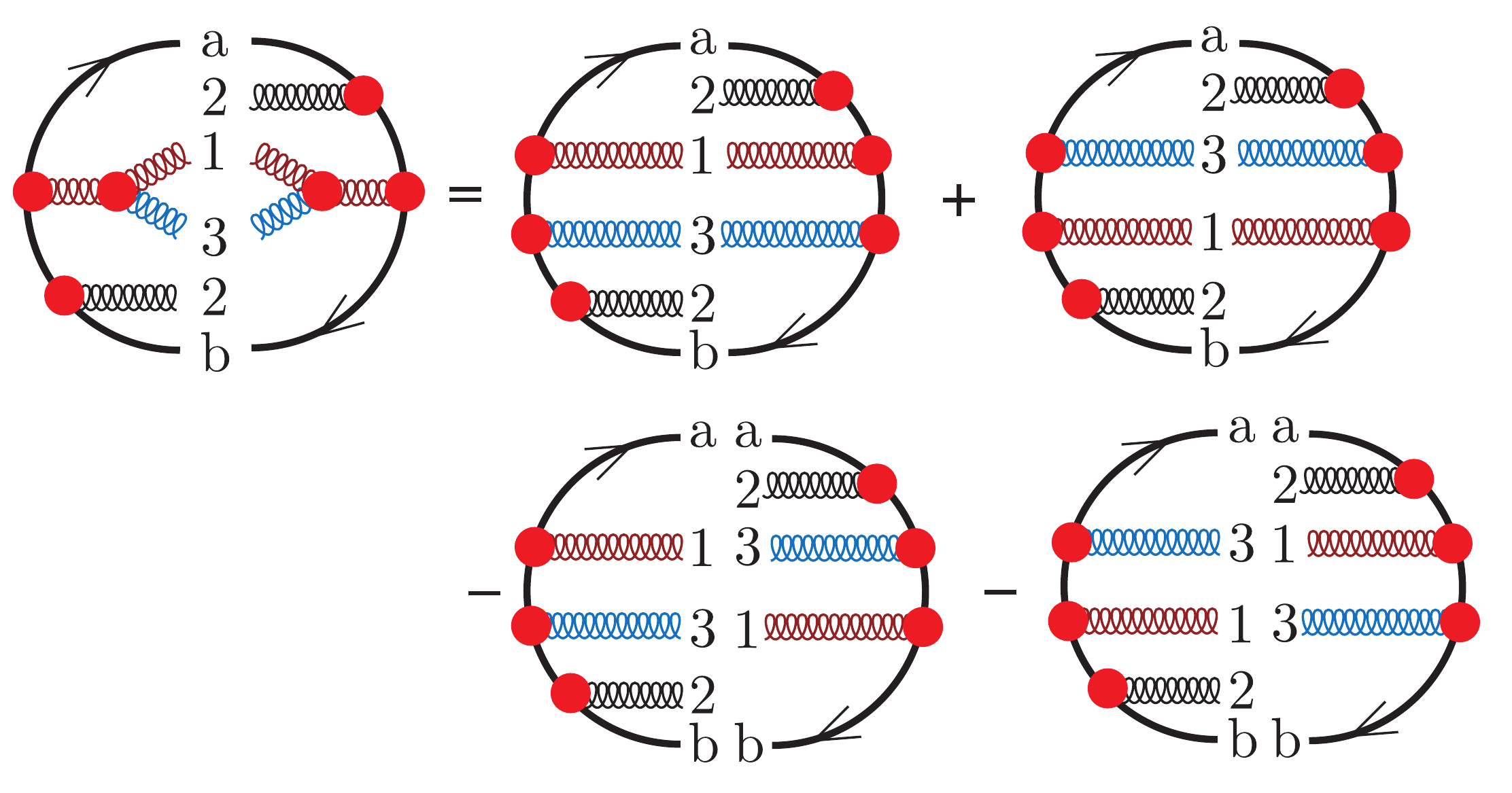}}

\caption{Identity for color dependence of splitting of gluon 1 in both the bra state and the ket state in the case $\{c'\}_m \ne \{c\}_m$. In the LC+ approximation, one keeps all four terms.  In the LC approximation, we keep only the first two terms.
}
\label{fig:splittingidentityketbra4}
\end{figure}

\begin{figure}
\centerline{\includegraphics[width = 14.4 cm]{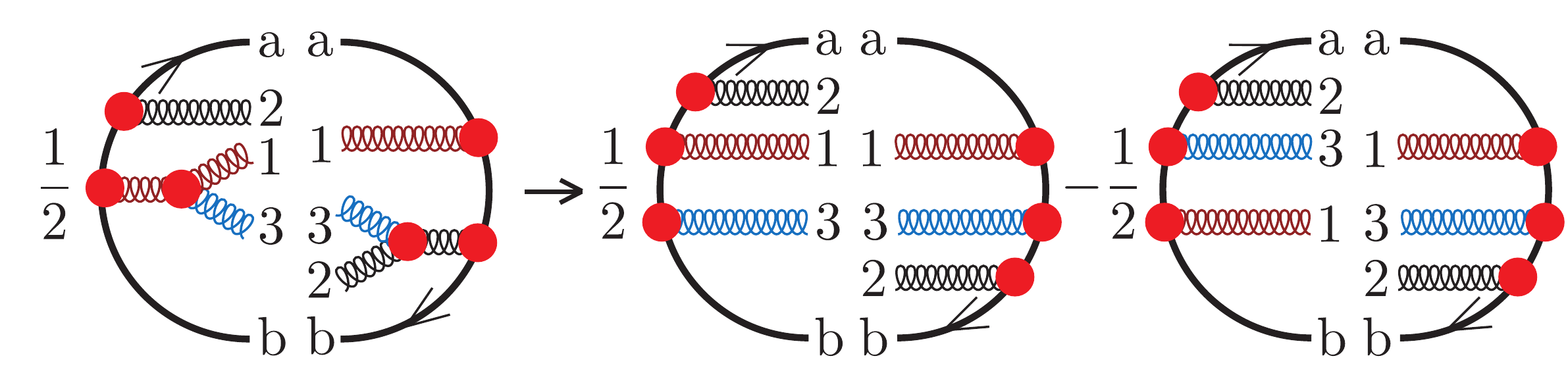}}

\caption{LC+ approximation for the splitting of gluon 1 in the ket state with the participation of helper parton 2 in the bra state in a case with $\{c'\}_m \ne \{c\}_m$. In the LC approximation, we keep only the first term.
}
\label{fig:splittingidentityketbra5}
\end{figure}

The evolution equation for the shower state as a function of the shower time $t$ has the form of a linear equation
\begin{equation}
\label{eq:evolution}
\frac{d}{dt}\sket{\rho(t)}
= [{\cal H}_I(t) - {\cal V}(t)]\sket{\rho(t)}
\;\;.
\end{equation}
Here ${\cal H}_I(t)$ is the parton splitting operator. It is given for full color in eq.~(5.7) of ref.~\cite{NScolor}. The LC+ approximation is defined in ref.~\cite{NScolor} by restricting the possibilities for color states that can be reached by a splitting. We can define the LC approximation by restricting these choices further. This is illustrated in figures \ref{fig:splittingidentityketbra4} and \ref{fig:splittingidentityketbra5}. 

In figure \ref{fig:splittingidentityketbra4} we illustrate the color structure for the splitting of a gluon into two gluons, numbered 1 and 3. The same splitting occurs in both the bra and ket states. Note that in this example, the color structures of the bra and ket states are not the same. Triple gluon vertices are $\mi f_{abc}$. Using eq.~(\ref{eq:f_abc}), we can expand the left hand side of figure \ref{fig:splittingidentityketbra4} into a linear combination of four products of color basis functions, as shown in the right hand side of figure \ref{fig:splittingidentityketbra4}. In the LC+ approximation, we keep all four of these terms. We define the LC approximation to keep just the first two terms. In the shower Monte Carlo algorithm, we choose one state or the other with probability 1/2.

In figure \ref{fig:splittingidentityketbra5} we illustrate the color structure for the radiation of a gluon 3 from gluon 1 in the ket state, but with interference with the radiation of gluon 3 from ``helper gluon'' 2 in the bra state. Expanding the left hand side into products of basis functions produces four terms. Of these, the LC+ approximation keeps the two terms shown in the right hand side of figure \ref{fig:splittingidentityketbra5}. We define the LC approximation to keep only the first term.

The general principle for emission of a gluon should be clear from these two examples: the new gluon (3) must not cross the emitting gluon (1) in the color diagram. For emission of a gluon from a quark, the situation is simpler: there is only one allowed structure in the LC+ approximation (as in figure 16 of ref.~\cite{NScolor}) and this structure is retained in the LC approximation. There is also the possibility of $\Lg \to q + \bar q$ splittings. With the color group $U(N_\Lc)$, there is only one possible new color state after the splitting. The new state is given by connecting the new quark and antiquark to the previous color string by using eq.~(\ref{eq:U3Fierz}).

We also need the operator ${\cal V}(t)$ in eq.~(\ref{eq:evolution}). This operator leaves the number of partons unchanged. Its exponential is the Sudakov factor that appears between splittings in the shower. We define ${\cal V}(t)$ from ${\cal H}_I(t)$ in such a way that the inclusive cross section that we started with at the hard interaction is not changed at all by the shower. The operator ${\cal V}(t)$ contains a color factor denoted by $N(k,l,\{\hat f\}_{m+1})$ in ref.~\cite{NScolor}. For the LC+ approximation, it is given in eq.~(6.12) of ref.~\cite{NScolor}. For the LC approximation, we need to calculate $N(k,l,\{\hat f\}_{m+1})$ using the color group $U(N_\Lc)$ and including only the LC contributions.  For all cases of gluon emission except for a $\Lg \to \Lg + \Lg$ splitting as in figure \ref{fig:splittingidentityketbra4}, $N(k,l,\{\hat f\}_{m+1})$ is $N_\Lc/2$, where the $N_\Lc$ comes from having one more ``quark'' loop and the 1/2 is the 1/2 in eq.~(\ref{eq:U3Fierz}). For a $\Lg \to \Lg + \Lg$ splitting, there are two terms, so the net color factor in ${\cal V}(t)$ is $2 (N_\Lc/2) = N_\Lc = C_\LA$. For a $\Lg \to q + \bar q$ splitting, $N(k,l,\{\hat f\}_{m+1})$ is just $T_\LR = 1/2$.

For the case of a $q \to q + g$ splitting in either the ket state or the bra state or both, the color factor in ${\cal V}(t)$ is $N_\Lc /2$ when we calculate this way. One normally uses $C_\LF$ instead. We arrange to obtain the conventional net factor of $C_\LF$ instead of $N_\Lc /2$ by simply multiplying the splitting probability by $2C_\LF/N_\Lc$ in ${\cal H}_I(t)$ for the LC approximation.

We need one more adjustment. When the change in color suppression index during the shower reaches $(\Delta I)_{\rm max}$, we switch from the LC+ approximation to the LC approximation as described above. This means changing the rules for calculating the color factor in quantum probabilities. If the color state before this switch is $\ket{\{c\}_m}\bra{\{c'\}_m}$ then the corresponding contribution to the inclusive cross section contains the color factor $\brax{\{c'\}_m}\ket{\{c\}_m}_{SU(N_\Lc)}$, where the subscript indicates a calculation using the color group $SU(N_\Lc)$. After the switch, we have the same nominal color state, $\ket{\{c\}_m}\bra{\{c'\}_m}$, but we have changed the definitions so that now the inclusive cross section contains the color factor $\brax{\{c'\}_m}\ket{\{c\}_m}_{U(N_\Lc)}$ calculated using the color group $U(N_\Lc)$. To preserve probabilities, we multiply the weight for the current shower state by
\begin{equation}
C_0 = \frac{\brax{\{c'\}_m}\ket{\{c\}_m}_{SU(N_\Lc)}}{\brax{\{c'\}_m}\ket{\{c\}_m}_{U(N_\Lc)}}
\;.
\end{equation}
This factor is normally close to 1, but in some situations the numerator can be zero. We note that \textsc{Deductor} contains fast algorithms for calculating both the numerator and the denominator.

\section{Another approach}
\label{sec:PlatzerSjodahl}

Pl\"atzer and Sj\"odahl have advocated another approach \cite{PlatzerSjodahl} to the problem of incorporating quantum color in a parton shower. This approach is dramatically different from that of refs.~\cite{NSI,NScolor}. 

Recall that for a parton shower with quantum color, the partonic state $\sket{\rho(t)}$ is a density matrix in the quantum color states. How ought $\sket{\rho(t)}$ to evolve in a parton shower with full color? In our opinion, the most basic requirement is the following. At the end of the shower at shower time $t_\Lf$, we have a partonic state $\sket{\rho(t_\Lf)}$ that can be expanded in powers of $\as$. The first order term in this expansion should be given by the Born approximation, $\sket{\rho_\LB}$, to $\sket{\rho}$ convoluted with collinear/soft approximations to one loop QCD graphs with an infrared cutoff corresponding to the final shower time $t_\Lf$. Ref.~\cite{PlatzerSjodahl} works with $\sket{\rho(t)}$, but the first order term in the expansion of $\sket{\rho(t_\Lf)}$ is not given by $\sket{\rho_\LB}$ convoluted with collinear/soft approximations to one loop QCD graphs with their correct color structure.

We can be more specific. Following the structure of quantum statistical mechanics and the idea that in a parton shower one parton can split into two partons, the evolution of the shower state can be represented as a linear equation, as in eqs.~(\ref{eq:rhofromU}) and (\ref{eq:evolutionU}) \cite{NSI}.  This equation involves an operator ${\cal H}_I(t)$ that increases the number of partons by one and an operator ${\cal V}(t)$ that leaves the number of partons and their momenta and flavors unchanged.  

Ref.~\cite{PlatzerSjodahl} has ${\cal H}_I(t)$ in agreement with ref.~\cite{NSI}. The color structure of ${\cal V}(t)$  is very well known \cite{NSI, EarlyGap1, EarlyGap2, Manchester2005, NonGlobal1, NonGlobal2, NonGlobal3, NonGlobal4, Manchester2009, SuperLeading1, SuperLeading2, Manchester2011, Platzer, Sjodahl}. However, in ref.~\cite{PlatzerSjodahl} with ``full color,'' the operator ${\cal V}(t)$ is absent.\footnote{It is noted in ref.~\cite{PlatzerSjodahl} that the authors would like to include ``virtual color rearranging gluon exchanges.'' We take this to mean that they would like to include ${\cal V}(t)$.} It is replaced by a Sudakov factor that does not change the partonic color state. The ``full color'' treatment of ref.~\cite{PlatzerSjodahl} is perhaps an approximation to a full color treatment, but it is not clear to us what sort of approximation it is or whether the approximation can be improved.


\end{document}